\documentclass{aa}

\usepackage{xcolor  }
\usepackage{txfonts }
\usepackage{natbib  }
\usepackage{hyperref}
\newcommand{\RA }[4]{$#1^\mathrm{h} #2^\mathrm{m} #3^\mathrm{s}    .#4$} % Right Ascension (J2000)
\newcommand{\Dec}[4]{$#1^\circ      #2^\prime     #3^{\prime\prime}.#4$} % Declination (J2000)
\newcommand{\N  }[1]{N_\mathrm{#1}}                                      % Column density of #1
\newcommand{\Jyb}{\textrm{Jy beam}$^{-1}$}  % rms noise:     Jy/beam
\newcommand{\am }{$^\prime$}                % angular dist.: arcminutes
\newcommand{\as }{$^{\prime\prime}$}        % angular dist.: arcsecs
\newcommand{\kms}{km~s$^{-1}$}              % velocity:      km/s
\newcommand{\cm }{cm$^{-3}$}                % density:       (particles)/cm^3
\newcommand{\um }{$\mu$m}                   % IR wavelength: micrometres
\newcommand{\CO  }{$^{12}$CO}               % molecular gas
\newcommand{\SNR }{G338.3$-$0.0}            % SNR's name
\newcommand{\HESS}{HESS~J1640$-$465}        % HESS source
\newcommand{\XMMU}{XMMU~J164045.4$-$463131} % XMMU source
\newcommand{\PSR }{PSR~J1640$-$4631}        % pulsar
\newcommand{\AGAL}{AGAL338.307$+$00.006}    % submillimetre clump
\begin{document}

\title{The environment of the $\gamma$-ray emitting SNR {\SNR}: a hadronic interpretation for {\HESS}}

\titlerunning{ISM around {\SNR}: a hadronic interpretation for {\HESS}}

\author{L.    Supan      \inst{1}
   \and A.~D. Supanitsky \inst{1}
   \and G.    Castelletti\inst{1}}

\institute{Instituto de Astronom\'ia y F\'isica del Espacio (IAFE, CONICET $-$ UBA) CC 67, Suc. 28, 1428 Buenos Aires, Argentina\\
\email{lsupan@iafe.uba.ar}}

\date{Received December XX, 2015 / Accepted Month XX, 2016}

\abstract
{The supernova remnant (SNR) {\SNR} spatially correlates with {\HESS}, which is considered the most 
luminous $\gamma$-ray source associated with a SNR in our Galaxy. The X-ray pulsar {\PSR} has been recently discovered within the SNR shell, which could favor a leptonic origin for the detected very-high-energy (VHE) emission. In spite of this, the origin of the VHE radiation from {\HESS} has not been unambiguously clarified so far. Indeed, a hadronic explanation cannot be ruled out by current observations. On the basis of atomic (HI) and molecular ({\CO}) archival data, we determine, for the first time, the total ambient density of protons in the region of the {\SNR}/{\HESS} system, a critical parameter for understanding the emission mechanisms at very high energies. The value obtained is in the $100-130$~{\cm} range.
Besides this, we developed a new hadronic model to describe the spectral energy distribution (SED) of the {\HESS} source, which includes the latest total $\gamma$-ray cross-section for proton-proton collisions available in the literature. By using the assessed ambient proton density, we found that the total energy in accelerated protons required 
to fit the data is $5.4^{+4.7}_{-2.3} \times 10^{49}\ \textrm{erg}$ and $1.6^{+1.4}_{-0.7} \times 10^{50}\ \textrm{erg}$ 
for a source distance of 8.5 and 13~kpc, respectively. The case where the source distance is 8.5~kpc agrees with the typical scenario in which the energy released is on the order of $10^{51}$~erg and $\sim10\%$ of that energy is transferred  
to the accelerated protons, whereas the case corresponding to a source distance of 
13~kpc requires either a higher value of the energy
released in the explosion or a larger energy fraction to accelerate protons.}


\maketitle

\section{Introduction}
\label{Introduction}

{\HESS} is considered the most luminous very-high-energy (VHE) $\gamma$-ray source associated with a Galactic supernova remnant (SNR). Since its discovery, it has been linked to the SNR {\SNR} because in the plane of the sky they appear spatially coincident 
\citep{Aharonian-05}. At TeV energies {\HESS} has been reported to be marginally extended with a Gaussian radius of 4{\am}.3 \citep{Abramowski-14-1640}, although the last statistical update on this source shows a more extended emission with a size of $\sim$12{\am} (see Fig. 1 in \citealt{Abramowski-14-1641}).\par
At radio wavelengths the remnant {\SNR} shows two bright arcs comprising an incomplete shell of $\sim$8{\am} in size, whose northern part is seen projected at the edge of a confusing area with several HII regions \citep{Castelletti-11}.
Recently, on the basis of {\sl NuSTAR} observations the previously known X-ray source {\XMMU} inside the radio shell of {\SNR}, was confirmed as a pulsar named \object{\PSR} (\citealt{Gotthelf-14} and references therein).\par
Roughly a decade after the discovery of {\HESS} \citep{Aharonian-05}, its origin is still an open question. Moreover, although the study of the medium in which {\SNR} evolves is crucial for understanding the origin of the VHE radiation, it has never been investigated in detail, and in particular, there is no observational determination of the ambient density reported to date.
In this work we analyze the environment of the SNR {\SNR} through a multiwavelength approach, and we use the observational data   to establish, for the first time, constraints on the ambient proton density.
Additionally, we investigate the plausibility of a hadronic mechanism for the $\gamma$-ray emission by modeling the GeV-TeV spectrum with the latest estimations available in the literature of the cross section for $\gamma$-ray production through proton-proton interactions.

\section{The data}
\label{Data}

We analyzed the surroundings of {\SNR} by using observations of both the atomic and the molecular gas. The neutral hydrogen (HI) data were extracted from the {\sl \emph{Southern Galactic Plane Survey}} (SGPS~I, \citealt{McClure-Griffiths-05}), which combines low- and high-resolution data from observations in the HI 21-cm line carried out with the 64-m single-dish Parkes telescope and the interferometer ATCA. The SGPS dataset has an angular resolution of 2{\am}.2 and a separation between consecutive velocity channels of 0.82~{\kms}. Radial velocities for this survey, as well as the ones used throughout this work, are measured with respect to the local standard of rest (LSR). The rms sensitivity of the HI survey is $\sim$1.6~K.
On the other hand, the molecular gas distribution was traced by using the emission corresponding to the $J=1-0$ rotational transition at 115 GHz of the {\CO}. 
These data were acquired with the 1.2-m telescope at the Cerro Tololo Inter-American Observatory in Chile, and they provide an angular resolution and a velocity channel separation of 0$^\circ$.125 and 1.30~{\kms}, respectively, with an rms noise per velocity channel of $\sim$0.12~K \citep{Dame-01}.\par

To trace the detailed radio continuum emission of {\SNR} and its surroundings we used the image at 610~MHz presented in \citet{Castelletti-11}. It was obtained from observations carried out with the Giant Metrewave Radio Telescope (GMRT) and has a synthesized beam of 12{\as}.6 $\times$ 5{\as}.0 and an rms noise level of 1.6 {m\Jyb}.\par

An image of the region around {\SNR} emitting at 870~{\um} was extracted from the APEX
{\sl  \emph{Telescope Large Area Survey of the Galaxy}} (ATLASGAL), carried out with the LABOCA bolometer array at the APEX telescope \citep{Schuller-09}. The angular resolution of the ATLASGAL image is $\sim$19{\as}.2, and the typical rms noise is $\sim$50$-$70~{m\Jyb}.\par

Additionally to the spectral line and continuum radio observations, three representative wavelengths were selected to reveal the far-infrared (far-IR) emission at 24, 160, and 250~{\um}. The map at the lowest wavelength was obtained from the 24 \emph{{\sl \emph{and}}} 70 {\sl \emph{Micron Survey of the Inner Galactic Disk with MIPS}} (MIPSGAL, \citealt{Carey-09}), carried out with the MIPS camera of {\sl Spitzer}.
The data at 160 and 250~{\um} acquired with the PACS \citep{Poglitsch-10} and SPIRE \citep{Griffin-10} instruments aboard {\sl Herschel}\footnote{{\sl Herschel} is an ESA space observatory with science instruments provided by European-led Principal Investigator consortia and with important participation from NASA.}, respectively, were retrieved from the {\sl Herschel Science Archive} (HSA).

\section{Results}
\label{Results}

\subsection{Observational constraints for the ISM around the SNR~{\SNR}}
\label{ISM}

\begin{figure}
  \centering
  \includegraphics[width=0.45\textwidth]{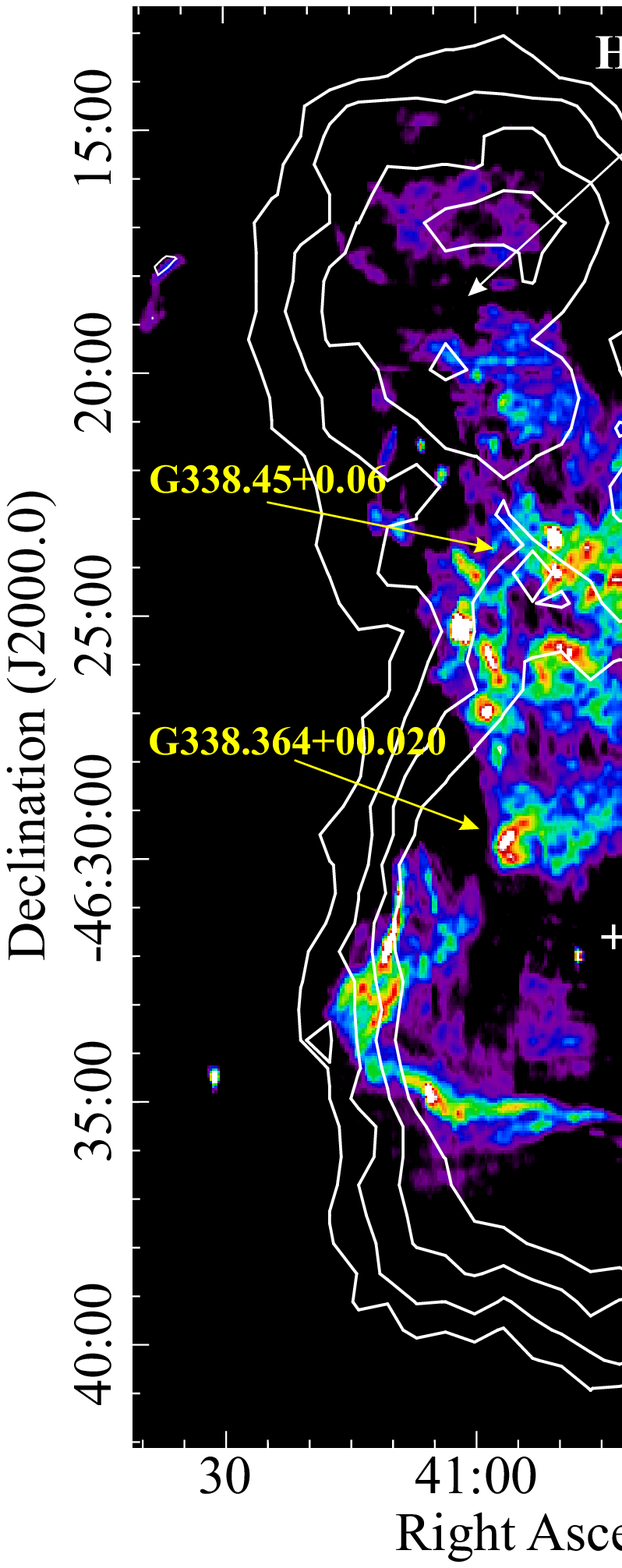}
  \caption{Field-of-view of {\SNR} as seen by the GMRT at 610~MHz with an angular resolution of 12{\as}.6 $\times$ 5{\as}.0 and a sensitivity of 1.6~{m\Jyb} \citep{Castelletti-11}. The linear scale is in units of {m\Jyb}. White contours delineate the VHE emission from {\HESS} with significance levels of 5, 6, 7, and $8\sigma$ \citep{Abramowski-14-1641}. The point spread function of the H.E.S.S. data is 0.$^\circ$085 (68\% containment region). The white cross indicates the position of {\PSR}. The three HII regions whose distance is reported by \citet{Urquhart-12} and the bubble G338.364$+$00.020 are also indicated.}
  \label{G338_610}
\end{figure}

Much of the discussion in the following depends on the radio continuum and $\gamma$-ray emission extents in the field toward {\SNR}. Therefore, before presenting our analysis of the medium in the vicinity of this remnant, we display in Fig.~\ref{G338_610} the GMRT high-resolution image at 610~MHz obtained in direction to {\SNR} overlaid with a few representative contours from the updated H.E.S.S. image of {\HESS} presented in \citet{Abramowski-14-1641}.
The bright region that appears to be in contact with the northern side of the SNR shell corresponds to the complex of thermal HII regions (\citealt{Urquhart-12}, \citealt{Brown-14}), whose names are also labeled in Fig.~\ref{G338_610} for reference.\par

A parameter of particular interest for determining the characteristics of the environment in which a SNR evolves is its distance. On the basis of HI 21-cm absorption and emission spectra, \citet{Lemiere-09} put {\SNR} at distances between 8.5 and 13~kpc, the same range that we adopt in our calculations. Although the connection between {\SNR} and the group of the northern HII regions has not been unambiguously proved, the general consistency of the SNR HI absorption spectrum with that obtained toward the ionized gas seems to support that they are physically related \citep{Lemiere-09}. This may convey the idea that {\SNR} is evolving in an inhomogeneous density environment.
Based on HI absorption profiles, \citet{Urquhart-12} established a distance of $\sim$ 13~kpc for the HII regions G338.39+0.16, G338.4+00.0, and G338.45+0.06 in the complex (see Fig. \ref{G338_610}).

\begin{figure}
  \centering
  \includegraphics[width=0.5\textwidth]{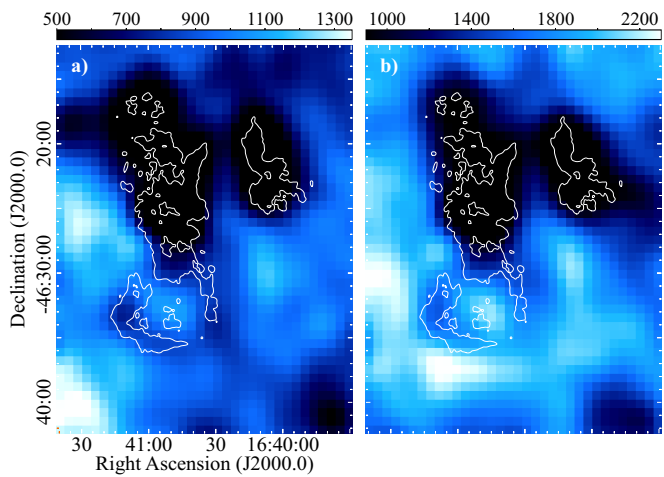}
  \caption{HI emission toward {\SNR} and the adjacent HII region complex, integrated in the velocity range {\bf a)} from $-121$ to $-111$~{\kms} and {\bf b)} from $-40$ to $-25$~{\kms}. The scales are linear in K~{\kms}. White contours delineate the radio emission at 610~MHz from Fig.~\ref{G338_610} at levels of 43 and 65 {m\Jyb}.}
  \label{HI_distribution}
\end{figure}

We study the ISM in the region of {\SNR} based on observations of the HI and {\CO} line intensities. Figure~\ref{HI_distribution} provides an overview of the atomic hydrogen emission integrated in the velocity intervals that correspond to the lower and upper kinematical distance limits to the SNR. Each image is the result of an integration in 16 consecutive spectral channels, spanning 13~{\kms} around the central velocity corresponding to distance values of 8.5 and 13~kpc (assuming the rotation curve for our Galaxy of \citealt{Fich-89}, with $R_0$ = 8.5~kpc and $\Theta_0$ = 220~{\kms}).
One can see that the HI gas depicts an inhomogeneous distribution, with remarkable minima exactly overlapping both the shell of {\SNR} and the ionized gas extending to the north. The remaining velocity images in the HI data cube show no features that may be clearly associated with {\SNR}.\par

\begin{figure}
  \centering
  \includegraphics[width=0.5\textwidth]{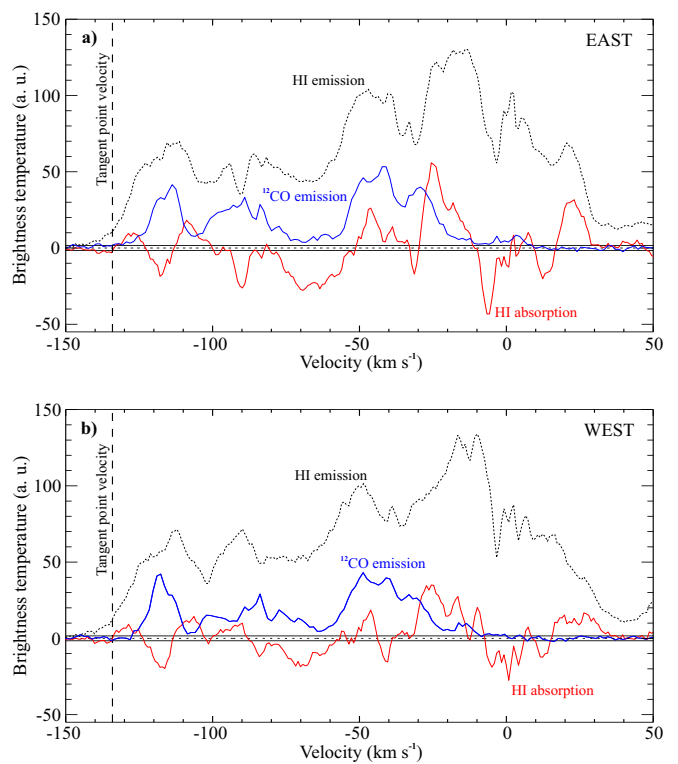}
  \caption{{\bf a)} {\CO} emission spectrum (blue line) and HI emission and absorption spectra (dashed black and red line, respectively) obtained in direction to the bright eastern rim of the {\SNR}. {\bf b)} The same as in {\bf a)}, toward the western rim of the remnant, where the source {\AGAL\ is located}.}
  \label{CO_arcs_spectra}
\end{figure}

Even though the lack of sufficient angular resolution of the {\CO} data prevents us from clearly distinguishing small-scale spatial correspondences between the shock front of {\SNR} and the molecular content in the region, we can still search for some kinematical evidence by analyzing the spectral properties of the interstellar gas. We thus construct combined spectra of the {\CO} emission and HI absorption in the direction of the radio-bright east and west limbs of the SNR.
Both spectra are presented in Fig.~\ref{CO_arcs_spectra}. The absorption profiles of the atomic gas, drawn by a red line in Fig.~\ref{CO_arcs_spectra}, were obtained after subtracting an average background spectrum to the emission one.
Several molecular components, represented by the blue line in both panels of Fig.~\ref{CO_arcs_spectra}, are also apparent in the velocity distribution of the {\CO}. Using the \citet{Fich-89} rotation curve, we found that the components at the highest negative velocities from $\sim$$-125$ to $\sim$$-107$~{\kms} (peak centered at $\sim$$-116$~{\kms}) and those in the velocity range between $\sim$$-40$ and $\sim$$-23$~{\kms} (peak centered at $\sim$$-31$~{\kms}), respectively, include the lower and upper limits set on the SNR distance.
In fact, owing to the kinematic distance ambiguity (KDA) in the fourth quadrant of the Galaxy, the peak centered at $\sim$$-116$~{\kms} corresponds to a distance of either 6.5~kpc or 9.3~kpc, whereas near and far distances of 2.6 and 13.2~kpc are deduced for the latter one centered at $\sim$$-31$~{\kms}.
To resolve the distance ambiguity, we note that in both spectra obtained toward the east and west rims, the {\CO} emission centered at $\sim$$-116$~{\kms} is associated with an absorption line in the HI 21-cm spectra up to the velocity of the tangent point.
The {\CO} emission seen around this velocity may, therefore, be the result of molecular material located beyond the tangent point, which is approximately $\sim$8~kpc. In this case, the corresponding HI absorption feature could be attributed to the atomic gas embedded within the {\CO}, which absorbs the continuum from {\SNR}.
In addition, spectral asymmetries toward red- and blueshifted velocities are evident at the highest negative velocity range for each SNR boundary.
Numerous factors can contribute to the observed broadening, among them perturbations in the velocity field of the molecular material caused by the interaction with the SNR shock front, thus providing an argument for the connection between the molecular emission and {\SNR}.
On the other hand, for both rims in {\SNR,} we consider the spectral match between the molecular emission line at $\sim$$-31$~{\kms} and possible absorption HI features somewhat speculative, and therefore it is difficult to reach any conclusion on the location of this molecular material.
However, as discussed below, the spatial correlation between the radio and the submillimeter continuum emission, and the sites emitting in the IR band, supports the hypothesis that these spectral features most probably reveal interstellar material in the surroundings of the SNR.

\begin{figure}
  \centering
  \includegraphics[width=0.5\textwidth]{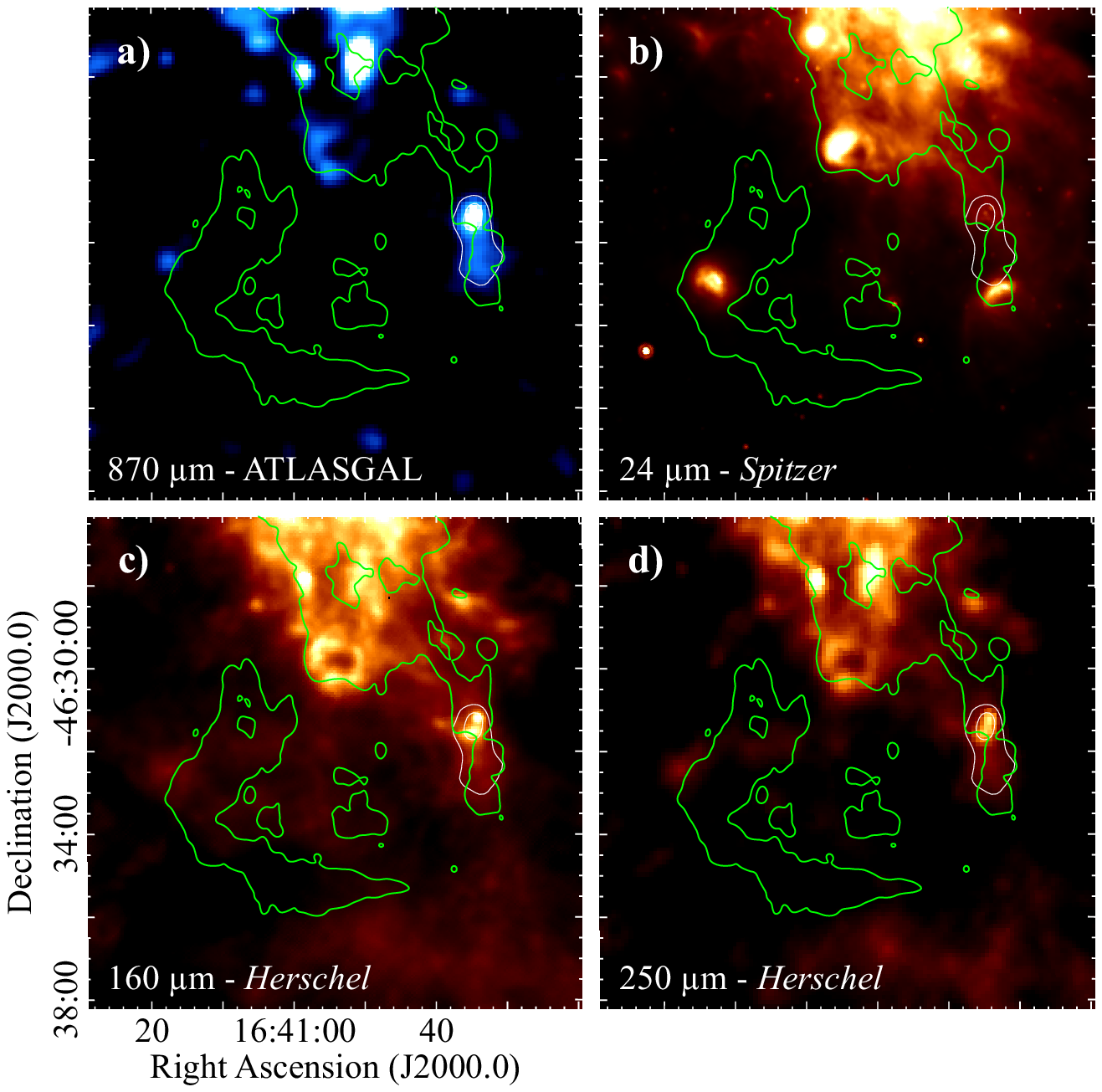}
  \caption{{\bf a)} 870-{\um} view of {\SNR} taken from the ATLASGAL. The source {\AGAL} is outlined by white contours. From {\bf b)} to {\bf d)}: far-IR emission in the field of {\SNR} at 24~{\um} from {\sl Spitzer}, and at 160 and 250~{\um} from {\sl Herschel}, respectively. The 4.3~m{\Jyb} green contour level from the 610~MHz image is included on each panel to facilitate the comparison between the submillimeter and IR emission with total power radio features.}
  \label{IR-AGAL_image}
\end{figure}

Supplementary evidence of the inhomogeneous medium that surrounds {\SNR} can be obtained from submillimeter and far-IR continuum emission across the field. Figure~\ref{IR-AGAL_image}a shows the 870~{\um} continuum emission as observed by APEX. In the middle part of the western radio-bright shell of {\SNR}, as seen in projection, there is a conspicuous bright clump (R.A. = \RA{16}{40}{34}{57}, Dec. = \Dec{-46}{31}{15}{9}, J2000.0) embedded in a diffuse emission region extending ($\sim$2{\am}) to the south. The angular extent of the clump, cataloged as the ATLASGAL source {\AGAL} \citep[CSC\footnote{CSC refers to the ATLASGAL - compact source catalog.}][]{Contreras-13}, is about 80{\as}. Panels b, c, and d in Fig.~\ref{IR-AGAL_image}, respectively, display the far-IR emission at 24~{\um} from {\sl Spitzer}, and in the two {\sl Herschel} bands of 160 and 250~{\um}. These images show that most of the long-wavelength IR emission is inside the intensity green contour that delineates the supernova shock front. Among the most remarkable IR features are the bubble G338.364$+$00.020 lying in the north side of the remnant \citep{Anderson-14} and the bright knot located in the western rim, likely the IR counterpart of the identified submillimeter clump. Additionally, a trail of diffuse IR emission is seen matching the western limb of {\SNR}. On the whole, the inhomogeneities observed from the atomic and molecular gas are thus supported by the morphological properties observed in the far-IR emission, which could arise from the interstellar dust embedded in the molecular material shocked and heated by the passage of the SNR shock front. 
Besides this, it is probably that the clump emitting at 870~{\um} corresponds to dense material within the molecular gas.\par

\subsection{The proton content in the SNR surroundings}
\label{Density}

\begin{figure}
  \centering
  \includegraphics[width=0.5\textwidth]{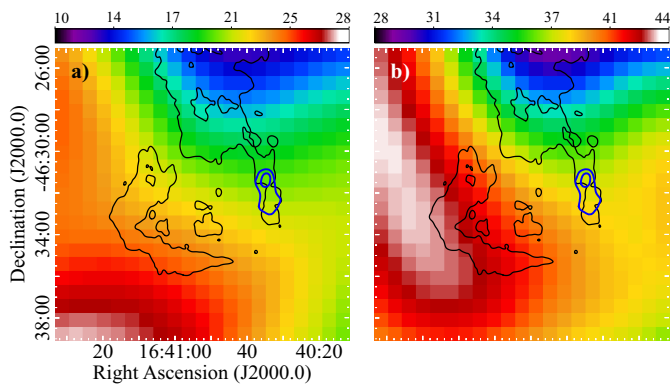}
  \caption{Total proton column density distribution toward {\SNR}, in the same velocity ranges as in Fig. \ref{HI_distribution}: {\bf a)} from $-121$ to $-111$~{\kms} and {\bf b)} from $-40$ to $-25$~{\kms}. The scale is in units of $10^{21}$~cm$^{-2}$. This image was obtained by summing up the proton contributions to the total column density from the atomic and molecular species of hydrogen. The resolution is the same as the {\CO} data. Black contours delineate the radio emission at 610~MHz at the same levels as in Fig. \ref{HI_distribution}. Blue contours indicate the source {\AGAL}.}
  \label{proton_distribution}
\end{figure}

In Fig.~\ref{proton_distribution} we present the distribution of the column density of the molecular and atomic protons in the region of {\SNR} for the velocity width $\Delta v$ = 13~{\kms} overlapping the velocities corresponding to the lower and upper kinematical distance determinations to {\SNR}. The superposed contours of the radio continuum emission at 610~MHz are plotted for reference. A gradient in the column density, hence in the distribution of the ISM, from the southeastern half of the field to the northwestern region dominates the picture.
For each interval of radial velocities $\Delta\nu$ considered in this work (see Fig. \ref{HI_distribution}), the hydrogen column density $\N{HI}$ was determined using the relation \citep{Dickey-90}
\begin{equation}
\label{HI_column}
  \N{HI} = 1.823 \times 10^{18} \int_{\Delta v} T_b \; dv, 
\end{equation}
which assumes that the emission is optically thin with a brightness temperature of the atomic gas emission represented by $T_b$.\par
An almost circular geometry defined by a mean radius of 0$^\circ$.12 was assumed for the zone in which we integrate the HI emission, according to the updated extension of {\HESS} \citep{Abramowski-14-1641}.
On the other hand, the proton contribution from the molecular hydrogen integrated in the same area was estimated through the column density of the molecular hydrogen $\N{H_2}$. It was inferred from the {\CO} dataset, through the integrated {\CO} emission $W_{^{12}\mathrm{CO}}$, using the {\CO}-to-H$_2$ conversion factor $X_\mathrm{CO} = \N{H_2}/W_{^{12}\mathrm{CO}} = 2.0 \times 10^{20}$~cm$^{-2}$ (K km s$^{-1}$)$^{-1}$
reported by \citet{Bertsch-93}. The total proton column density is therefore obtained taking both the atomic and molecular contributions into account: $\N{p} = \N{HI} + 2\N{H_2}$. For the highest velocity range, the integration from $-121$ to $-111$~{\kms} yields a total column density $\N{p} \sim 6.6 \times 10^{21}$ cm$^{-2}$, while for the interval between $-40$ and $-25$~{\kms}, the column is $\N{p} \sim 8.7 \times 10^{21}$ cm$^{-2}$. In both cases the main contribution to $\N{p}$ is given by the molecular hydrogen. We then use the derived column density to roughly calculate the average proton density $\bar{n}_\mathrm{p}=\N{p}/L$ for the TeV region, where we assume that the thickness $L$ along the line of sight equals the size of the minor axis of the integration region ($L \sim 36$~pc and $L \sim 55$~pc at distances of 8.5 and 13~kpc, respectively). The total proton density estimated in this way varies between 100 and 130~{\cm} for the upper and lower distance limits, respectively, with uncertainties of about 30\%. The associated total proton masses are approximately $2\times10^5$ and $4\times10^4$ $M_\odot$ for each density value.\par
Additionally, we infer the mass and density of the gas in the bright clump identified at submillimeter wavelengths, using the relation given by \citet{Abreu-Vicente-15} for ATLASGAL sources
\begin{equation}
\label{AGAL_mass}
  M = \frac{S_\nu \; R \; d^2}{B_\nu(T_\mathrm{d}) \; \kappa_\nu},
\end{equation}
where $\nu$ represents a frequency of 345~GHz (corresponding to 870~{\um}) at which an integrated flux density $S_\nu$ is measured over the source, $d$ is the distance to the clump, and $B_\nu(T_\mathrm{d})$ is the Planck distribution function at the dust temperature $T_\mathrm{d}$, assumed to be $\sim$20~K \citep{Schuller-09}. Here, $R$ is the gas-to-dust ratio, and $\kappa_\nu$ is the dust absorption coefficient. If we adopt canonical values of 100 and $1.85~\mathrm{cm}^2~\mathrm{g}^{-1}$ for $R$ and $\kappa_\nu$ \citep{Schuller-09} and use the flux density measurement for {\AGAL} reported in the CSC catalog $S_\nu$ = (8.48~$\pm$~1.55)~Jy (which corresponds to a region with an effective radius of about 40{\as}), we obtain a total mass of $\sim$$4.2\times10^3M_\odot$. In this calculation we assume an average distance of $\sim$10~kpc to the clump.
The hydrogen particle density of the clump can be calculated through the relation \citep{Rosolowsky-10}
\begin{equation}
\label{AGAL_density}
  n_\mathrm{H} = \frac{3M}{4 \pi r^3 \mu_\mathrm{H} m_\mathrm{H}},
\end{equation}
where $r$ is the radius of the ATLASGAL source, while $\mu_\mathrm{H}=2.8$ and $m_\mathrm{H}$ are the mean particle mass and the hydrogen atom mass, respectively. The resulting density is $\sim$$2.5\times10^3$~{\cm}. Both the mass and density obtained here for {\AGAL} are comparable to those found in other ATLASGAL sources with similar characteristics to the clump analyzed here \citep{Schuller-09}.

\section{$\gamma$-ray  production in the SNR G338.3-0.0/HESS J1640-465 system}
\subsection{Background on $\gamma$-ray emission models}
\label{Background}

To put the calculations presented in this work in context, we first
briefly review the models that over the past eight years have been proposed to account for the spectral energy distribution (SED) of the {\SNR}/{\HESS} system. At the beginning, the detection of the source {\XMMU} \citep{Funk-07}, which was by then a promising pulsar candidate inside the radio extent of {\SNR}, made the pure leptonic processes the most natural way to explain the observed TeV $\gamma$ rays. Indeed, a variety of leptonic models have been elaborated, in which the $\gamma$-ray photons are primarily created via inverse Compton (IC) interactions of energetic electrons injected by the compact source that scatter with ambient photons \citep{Lemiere-09,Slane-10}.

Later on  to model the flux measurements from
radio to the $\gamma$-ray band, \citet{Abramowski-14-1640} included the contribution to the TeV photons resulting from interactions of electrons accelerated at the SNR shock front with their surroundings.
However, as noted by the authors themselves, their approach with fixed IC contribution from the dust content and electron/proton ratio fails in providing a good fit to the $\gamma$-ray spectral shape. A possible contribution by electron bremsstrahlung emission was also considered by \citet{Abramowski-14-1640} to explain the origin of {\HESS}, but they found that a high target density of 500~{\cm} and high electron/proton ratios of about 0.1 are needed to match the TeV fluxes of the source. 

On the basis of the physical parameters derived for the recently discovered pulsar {\PSR} within the SNR~{\SNR}, \citet{Gotthelf-14} revised an evolutionary model for the pulsar wind nebula (PWN) in which the broad-band spectrum for {\HESS} is fit with a single power-law function. With a pulsar distance assumed to be 12~kpc, 
\citet{Gotthelf-14} find a fitted low ambient density of 0.03~{\cm}.
It is interesting to point out that in all the aforementioned leptonic models,
the radio band is poorly sampled. Indeed, except in the \citet{Abramowski-14-1640} work, in which the radio flux estimates at 235, 610, 1280, and 2300~MHz \citep{Castelletti-11} were arbitrarily scaled (by a factor of 0.5), in the remaining works only the flux upper limit determined at 610~MHz \citep{Giacani-08} or the flux at 843~MHz \citep{Whiteoak-96} were considered to anchor the spectrum of {\HESS} at the lower energy end.

Different hadronic interaction models were also applied to account for the spectral distribution of {\HESS}. For the hadronic emission, the observed $\gamma$ rays would be generated in the decay of neutral pions created after the collisions of protons accelerated at the SNR shock with dense interstellar material.
However, in all the works that analyze this alternative scenario, no observational constraints on the surrounding medium density were used, but this relevant magnitude was either derived from the fitted 
$\gamma$-ray spectrum or constrained during the modeling. For instance, low ambient densities of $\sim$1~{\cm}, for distances less than 8~kpc, and the SN total energy required to accelerate hadrons 
as large as $\sim 7.6 \times 10^{50}$~erg were deduced in the model proposed by \citet{Funk-07}. On the other hand, \citet{Slane-10} find that hadronic modeling is possible with an assumed shock compression ratio of 4 and a high total SN energy transferred to accelerated protons of 25~\%, which fix the ISM density at $\sim$100~{\cm}. 

Similar values for the ambient density were deduced by \citet{Abramowski-14-1640}.
However, they determined that the product of total proton energy and mean target density could be as high as $W_\mathrm{p} \, \bar{n}_\mathrm{H} \sim 4 \times 10^{52}$~erg~{\cm}, assuming a distance of 10~kpc to {\SNR}. A more recent estimation of the high-energy part of the {\HESS} SED has been reported in \citet{Lemoine-Goumard-14}, using the latest {\sl Fermi}-LAT data. In that work a harder SED compared with earlier studies was obtained. These authors also developed a hadronic model to explain both the high- and very-high-energy part of the SED by using the new {\sl Fermi}-LAT fluxes. Based on the previous considerations from 
\citet{Abramowski-14-1640}, they used an ambient proton density of 150~{\cm} and concluded that a hadronic scenario is a viable option. However, they found difficulties in discarding a component originated in the PWN, which could relax the constrain of high values of energy in accelerated protons that require the hadronic scenario in some regions of the parameter space.

In the following, we show that the GeV-TeV $\gamma$-ray spectrum of {\HESS}, which incorporates the latest data collected with the {\sl Fermi}-LAT and H.E.S.S. instruments, can be explained on the basis of the pure hadronic model developed in the current work. Our new analysis uses target proton densities measured in the region of the $\gamma$-ray source to conciliate the difficulties of earlier models by bringing together reasonable values for the SN explosion energy and the efficiency of converting this energy into shock accelerated protons.

\subsection{An improved hadronic model}
\label{SED}

As already discussed in Sec.~\ref{Background}, a hadronic origin of the $\gamma$-ray emission from {\HESS} has been previously proposed in the literature. In this section we present a new model to account for the production of $\gamma$ rays in the HESS source via hadronic interactions. 
As an important particularity, our emission model uses the parametrization of the $\gamma$-ray differential cross section in the proton-proton 
collisions given in \citet{Kafexhiu-14}, which is more detailed and comprehensive than 
previously published ones \citep{Kelner06,Kamae06}.
Consequently, the differential cross-section,
especially in the low proton energy part, together with the fit of the spectral index and the cutoff energy (which are fixed a priori in previous 
models) are now described much better. In modeling the spectra, we consider the latest statistics data from {\sl Fermi}-LAT and H.E.S.S. of {\HESS}. 
The total energy going into accelerated protons required for fitting all the collected data points, which is an essential physical parameter to test the
feasibility of the model, is assessed by using the proton density in the region of the $\gamma$-ray emission obtained in Sec.~\ref{ISM},  
$\bar{n}_\mathrm{p} \simeq 100-130$~{\cm}. It is worth mentioning that the relevance of this magnitude in our model was the main motivation of a 
reliable determination of the ambient protons density in the region of the source.

We fit the $\gamma$-ray data points by assuming a proton distribution in momentum given by a power law with an exponential cutoff

\begin{equation}
\label{eqn_prspec}
  \frac{dN}{dp} = C\ p^{-\Gamma} \exp \left(-p/p_\mathrm{cut} \right),
\end{equation}

\noindent
where $C$ is a normalization constant; $p$  the momentum of the accelerated protons, which is related to the proton energy by $E_p=\sqrt{\smash[b]{c^2 p^2 + m_p^2 c^4}}$ ($c$ is the speed of light and $m_p$  the proton mass), $\Gamma$ is the spectral index, and $p_\mathrm{cut}$ represents the cutoff momentum. The cutoff proton energy is given by $E_\mathrm{cut}=\sqrt{\smash[b]{{c^2} p^2_\mathrm{cut} + m_p^2 c^4}}$.\par
The $\gamma$-ray flux at Earth can be written as

\begin{equation}
\label{eqn_gflux}
  J(E_\gamma) = \frac{c}{4\pi\ d^2}\ \bar{n}_\mathrm{p} \int_{E_\gamma}^\infty dE_p \; \frac{dN}{dE_p}(E_p) %
                \; \frac{d\sigma}{dE_\gamma}(E_\gamma,E_p),
\end{equation}

\noindent
where $d$ is the distance to the source and $d\sigma/dE_\gamma$ is the differential cross-section in proton-proton collisions resulting in the $\gamma$-ray emission. The weighted proton density in the region of the source is given by

\begin{equation}
\label{eqn_density}
  \bar{n}_\mathrm{p} = \int_V d^3x \; n_\mathrm{p}(\mathbf{x}) \; P_\mathrm{p}(\mathbf{x}),
\end{equation}

\noindent
where $V$ is the volume of the source, $P_\mathrm{p}(\mathbf{x})$  the spatial distribution function of the accelerated protons, and $n_\mathrm{p}$  the total ambient density. It should be noted that for the case where the accelerated protons are uniformly distributed in the source volume $P_\mathrm{p}(\mathbf{x})=1/V$, so $\bar{n}_\mathrm{p}$ is the average of the proton density in the source volume. We assume that for all subsequent calculations.

The energy spectrum of $\gamma$ rays originated in proton-proton interactions has been extensively studied in the literature. Recently, a detailed 
study has been performed by \citet{Kafexhiu-14}. In that work, a parametrization of the $\gamma$-ray energy spectrum has been obtained for proton energies ranging from 
the proton-proton kinematic threshold to 1~PeV. The differential cross section at low proton energies 
($\lesssim 3$~GeV) is obtained from a compilation of experimental data, while at high proton energies it is derived from Monte Carlo simulations. The hadronic interactions 
at the highest energies are not known. In particular, it is not possible yet to predict the results of these interactions from first principles 
quantum chromodynamics \citep[see for instance,][]{Werner93}. Nevertheless, there are models that make use of low energy accelerator data to extrapolate 
the hadronic interactions to higher energies \citep{Enterria11}. The parametrization in \citet{Kafexhiu-14} has been performed for the hadronic interaction models
implemented in Geant 4.10.0 \citep{geant4}, PYTHIA 8.18 \citep{pythia}, Sibyll 2.1 \citep{sibyll}, and QGSJET01 \citep{qg01}. 
In this paper, 
we use the PYTHIA 8.18 option for the high-energy part.
It is worth mentioning
that one of the most important aspects of the new parametrization is the detailed description of the $\gamma$-ray spectrum at low proton energies, 
which represents an important improvement over older approaches. 

\begin{figure}
  \centering
  \includegraphics[width=0.5\textwidth]{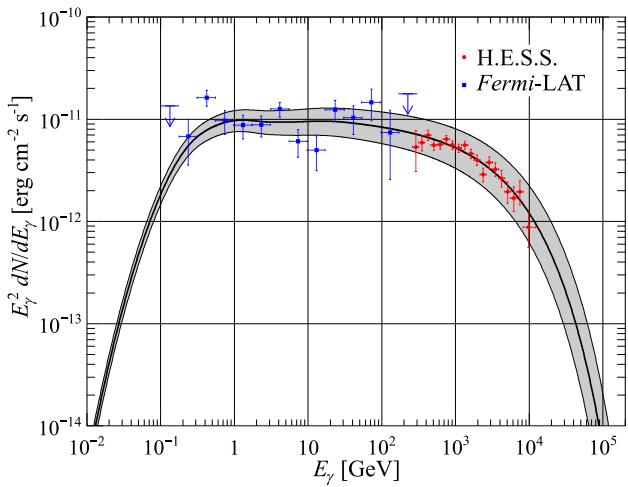}
  \caption{Spectral energy distribution of {\HESS} in the $\gamma$-ray band. The squares correspond to \emph{Fermi}-LAT data, while filled circles correspond to H.E.S.S. observations. The solid line is a $\chi^2$ fit of the SED, and the shaded area is the 1$\sigma$ region of that fit.}
  \label{SFit}
\end{figure}

In Fig.~\ref{SFit} we show the SED of {\HESS} for high- and very-high energies. The data points of the high-energy part of the SED correspond to observations carried out by {\sl Fermi}-LAT \citep{Lemoine-Goumard-14}, whereas the measurements are from H.E.S.S. in the
VHE region \citep{Abramowski-14-1640}. We performed a $\chi^2$-fitting to all the $\gamma$-ray points. The solid line included in Fig.~\ref{SFit} denotes our fit, obtained by using the Eqs.~(\ref{eqn_prspec}) and (\ref{eqn_gflux}) presented here and leaving $C$, $\Gamma$, and $E_\mathrm{cut}$ as free parameters. 
Our fit provides a spectral index $\Gamma = 2.13 \pm 0.02$ and a cutoff energy log($E_\mathrm{cut}$/GeV)= 4.73~$\pm$~0.09.\par

From Fig.~\ref{SFit} it can be seen that the $\gamma$-ray spectral shape is quite well fitted by our model, with a $\chi^2 = 38.1$ for 28 {\it d.o.f}. As mentioned above, it is mandatory to calculate the total energy in accelerated protons required by the model to fit the data points, in order to test the feasibility of it. The total energy in accelerated protons is given by

\begin{equation}
\label{eqn_PEn}
  \mathcal{E}_\mathrm{p} = \int_{E_\mathrm{min}}^\infty dE_p \; E_p \; \frac{dN}{dE_p}(E_p),
\end{equation}

\noindent
where it is assumed that $E_\mathrm{min} = 1.218$~GeV, which is the threshold energy for pion production \citep{Kafexhiu-14}. In this way the minimum energy in accelerated protons is calculated, i.e., accelerated protons with lower energies could exist but they cannot produce pions when interacting with ambient protons, and then they do not contribute to the $\gamma$-ray spectrum.\par

For the case in which {\HESS} is placed at the lower distance limit of 8.5~kpc, the value determined for the SNR~{\SNR}, the total energy in accelerated protons is calculated by using Eq.~(\ref{eqn_PEn}) taking into account the uncertainty on the ambient proton density and on the flux at 1$\sigma$ level (see shaded region in Fig.~\ref{SFit})

\begin{equation}
  \mathcal{E}_\mathrm{p} = \left(5.4^{+4.7}_{-2.3} \right)\times 10^{49}\ \textrm{erg} \ \ \ \ (d=8.5\ \textrm{kpc}),
\end{equation} 

\noindent
in this expression the central value is obtained by using the best fit to the $\gamma$-ray spectral data points and a proton density $\bar{n}_\mathrm{p}=130$ cm$^{-3}$. On the other hand, for the upper limit of 13~kpc, we obtain the following energy range

\begin{equation}
  \mathcal{E}_\mathrm{p} = \left(1.6^{+1.4}_{-0.7} \right)\times 10^{50}\ \textrm{erg}\ \ \ \ (d=13\ \textrm{kpc}),
\end{equation}

\noindent
where the central value is obtained by using the best fit to the $\gamma$-ray spectrum and a density $\bar{n}_\mathrm{p}=100$~{\cm}.\par

The typical energy released in a SN explosion is assumed to be on the order of $10^{51}$~erg, and the canonical fraction of energy transferred to the accelerated particles is 10\%. In the case where the source is placed at $d=8.5$~kpc, the total energy in accelerated protons is 
$\lesssim 10^{50}$~erg, which is consistent with the typical scenario.
However, when the source is placed at $d=13$~kpc, the total energy in accelerated protons can be as high as $\sim 3 \times 10^{50}$~erg, which is three times greater than the typical value. We note that $\mathcal{E}_\mathrm{p} \sim 10^{50}$~erg is the lower end of the interval. In any case, higher values for the total energy in accelerated protons could be due to either more energy being released in the supernova explosion or a larger amount of energy transferred to this component. Another possibility is the existence of an additional contribution to the $\gamma$-ray spectrum because of the PWN in G338.3$-$0.0 as suggested by \citet{Lemoine-Goumard-14}.\par

As mentioned in Sec.~\ref{Density}, the source {\AGAL} can be a prominent region of $\gamma$-ray emission due to its high value of proton density compared with the average one. The fraction of the total $\gamma$-ray flux coming from this region can be estimated through

\begin{equation}
  f_\mathrm{clump} = \frac{J_\mathrm{clump}}{J} = \frac{\bar{n}_\mathrm{p}^\mathrm{clump} V_\mathrm{clump}}{\bar{n}_\mathrm{p}\ V},
\end{equation}  

\noindent
where $\bar{n}_\mathrm{p}^{\mathrm{clump}}$ and $V_{\mathrm{clump}}$ are the average density and volume of {\AGAL},  while $\bar{n}_\mathrm{p}$ and $V$ are the average density and volume of the entire $\gamma$-ray source, respectively.
The fraction of the flux for a distance $d=8.5$~kpc is $f_{\mathrm{clump}} \cong 1.6 \%$, which was obtained using that $\bar{n}_\mathrm{p}^{\mathrm{clump}}=3\times10^3$~cm$^{-3}$ (calculated from Eq.~(\ref{AGAL_density})) and $\bar{n}_\mathrm{p}=130$~cm$^{-3}$.
The corresponding value of this fraction for $d=13$~kpc is $f_{\mathrm{clump}} \cong 1.2 \%$, computed with $\bar{n}_\mathrm{p}^{\mathrm{clump}}=1.8\times10^3$~cm$^{-3}$ (also from Eq.~(\ref{AGAL_density})) and $\bar{n}_\mathrm{p}=100$~cm$^{-3}$. The volume of both regions depends on the source distance. The cataloged angular size is used to calculate the volume of {\AGAL}. 
For the source {\HESS,} we use the volume corresponding to the calculation of the average density (see Sec.~\ref{Density}). In any case, it seems that the {\AGAL} region makes a subdominant contribution to the total $\gamma$-ray emission of the source.

\section{Summary}
\label{Summary}

In this paper, based on a multiwavelength approach, we have analyzed the morphological and spectral properties of the ISM in the vicinity of 
the SNR~{\SNR}. Taking the spatial match between the $\gamma$-ray source {\HESS} and the entire shell of the remnant into account, we revisited 
the likelihood of a hadronic nature for the TeV photons. For this purpose, we analyzed the HI, {\CO}, sub-mm continuum, and far-IR available 
information and showed that this remnant evolves in a highly inhomogeneous ambient with strong gradients in the distribution of the atomic and 
molecular gas. We have also found possible observational signatures of a physical relationship between the SNR and the {\CO} material.
Additionally, for the first time, we used the atomic and molecular observational
lines to determine the averaged ISM proton density in the $100-130$~{\cm} range over the whole extension of {\HESS} 
(and hence the SNR shell).\par

Our findings, obtained directly from observational data in the region of the SNR~{\SNR}/{\HESS} system rather than from the fitting models as 
occurred in all the previous studies provide support to show that the $\gamma$-ray emissivity can be suitably modeled by a purely hadronic 
scenario. Moreover, the assessed target proton density allowed us to study the feasibility of our model by comparing the energy in accelerated
protons, required to fit the $\gamma$-ray {\HESS} spectrum, with the typical scenario in which the energy released by the SN explosion is on the order 
of $10^{51}$~erg, and approximately 10~\% of the shock kinetic energy goes to the relativistic particles. For the case where the system is at a 
distance of 8.5~kpc, the energy in accelerated protons is consistent with the typical scenario, but when the source is at a distance 
of 13~kpc, the energy in accelerated protons can be larger (up to a factor of three). The latter case can be explained by the higher energy released 
during the SN explosion or even by a larger energy fraction transferred to the accelerated protons. In either case, a contribution from the PWN 
associated with the pulsar {\PSR} cannot be discarded by the present data. The next generation of Cherenkov telescopes like the Cherenkov Telescope 
Array (CTA, \citealt{Actis-11}) will have a much better angular resolution, which will be very useful for identifying the regions where 
most of $\gamma$ rays originate. This is of crucial importance for studying the relative contributions of the PWN and the {\SNR} and surroundings regions.

It is worth mentioning that the average proton density estimated in our work should not have a strong impact on the hadronic models found in 
the literature \citep[like the one in][]{Abramowski-14-1640} because the proton density required or assumed in most of these models is $\sim150$~cm$^{-3}$, a value contained in the interval of densities obtained in our work for a source distance of 8.5~kpc.

\begin{acknowledgements}
The authors acknowledge the anonymous referee's comments that helped to improve the article. A.~D. Supanitsky and G. Castelletti are members of the {\it Ca\-rre\-ra del Investigador Cient\'ifico} of CONICET, Argentina. L. Supan is a PhD Fellow of CONICET, Argentina.
This research was supported by CONICET (Argentina) through the Grant PIP 360/11 and by ANPCyT (Argentina) through the Grants PICT 902/13 and 0571/11.
This research has made use of the NASA/IPAC Infrared Science Archive, which is operated by the Jet Propulsion Laboratory, California Institute of Technology, under contract with the National Aeronautics and Space Administration.
The ATLASGAL project is a collaboration between the Max-Planck-Gesellschaft, the European Southern Observatory (ESO) and the Universidad de Chile. It includes projects E-181.C-0885, E-078.F-9040(A), M-079.C-9501(A), M-081.C-9501(A), plus Chilean data.
\end{acknowledgements}

\bibliographystyle{aa}
%\bibliography{supan-aa27962}
\bibliography{aa27962}

\begin{thebibliography}{36}
\expandafter\ifx\csname natexlab\endcsname\relax\def\natexlab#1{#1}\fi

\bibitem[{{Abramowski} {et~al.}(2014{\natexlab{a}}){Abramowski}, {Aharonian},
  {Ait Benkhali}, {Akhperjanian}, {Ang{\"u}ner}, {Backes}, {Balenderan},
  {Balzer}, {Barnacka}, {Becherini}, \& et~al.}]{Abramowski-14-1641}
{Abramowski}, A., {Aharonian}, F., {Ait Benkhali}, F., {et~al.}
  2014{\natexlab{a}}, \apjl, 794, L1

\bibitem[{{Abramowski} {et~al.}(2014{\natexlab{b}}){Abramowski}, {Aharonian},
  {Benkhali}, {Akhperjanian}, {Ang{\"u}ner}, {Anton}, {Balenderan}, {Balzer},
  {Barnacka}, {Becherini}, \& et~al.}]{Abramowski-14-1640}
{Abramowski}, A., {Aharonian}, F., {Benkhali}, F.~A., {et~al.}
  2014{\natexlab{b}}, \mnras, 439, 2828

\bibitem[{{Abreu-Vicente} {et~al.}(2015){Abreu-Vicente}, {Kainulainen},
  {Stutz}, {Henning}, \& {Beuther}}]{Abreu-Vicente-15}
{Abreu-Vicente}, J., {Kainulainen}, J., {Stutz}, A., {Henning}, T., \&
  {Beuther}, H. 2015, \aap, 581, A74

\bibitem[{{Actis} {et~al.}(2011){Actis}, {Agnetta}, {Aharonian},
  {Akhperjanian}, {Aleksi{\'c}}, {Aliu}, {Allan}, {Allekotte}, {Antico},
  {Antonelli}, \& et~al.}]{Actis-11}
{Actis}, M., {Agnetta}, G., {Aharonian}, F., {et~al.} 2011, Experimental
  Astronomy, 32, 193

\bibitem[{{Agostinelli} {et~al.}(2003){Agostinelli}, {Allison}, {Amako},
  {Apostolakis}, {Araujo}, {Arce}, {Asai}, {Axen}, {Banerjee}, {Barrand},
  {Behner}, {Bellagamba}, {Boudreau}, {Broglia}, {Brunengo}, {Burkhardt},
  {Chauvie}, {Chuma}, {Chytracek}, {Cooperman}, {Cosmo}, {Degtyarenko},
  {Dell'Acqua}, {Depaola}, {Dietrich}, {Enami}, {Feliciello}, {Ferguson},
  {Fesefeldt}, {Folger}, {Foppiano}, {Forti}, {Garelli}, {Giani},
  {Giannitrapani}, {Gibin}, {G{\'o}mez Cadenas}, {Gonz{\'a}lez}, {Gracia
  Abril}, {Greeniaus}, {Greiner}, {Grichine}, {Grossheim}, {Guatelli},
  {Gumplinger}, {Hamatsu}, {Hashimoto}, {Hasui}, {Heikkinen}, {Howard},
  {Ivanchenko}, {Johnson}, {Jones}, {Kallenbach}, {Kanaya}, {Kawabata},
  {Kawabata}, {Kawaguti}, {Kelner}, {Kent}, {Kimura}, {Kodama}, {Kokoulin},
  {Kossov}, {Kurashige}, {Lamanna}, {Lamp{\'e}n}, {Lara}, {Lefebure}, {Lei},
  {Liendl}, {Lockman}, {Longo}, {Magni}, {Maire}, {Medernach}, {Minamimoto},
  {Mora de Freitas}, {Morita}, {Murakami}, {Nagamatu}, {Nartallo}, {Nieminen},
  {Nishimura}, {Ohtsubo}, {Okamura}, {O'Neale}, {Oohata}, {Paech}, {Perl},
  {Pfeiffer}, {Pia}, {Ranjard}, {Rybin}, {Sadilov}, {Di Salvo}, {Santin},
  {Sasaki}, {Savvas}, {Sawada}, {Scherer}, {Sei}, {Sirotenko}, {Smith},
  {Starkov}, {Stoecker}, {Sulkimo}, {Takahata}, {Tanaka}, {Tcherniaev}, {Safai
  Tehrani}, {Tropeano}, {Truscott}, {Uno}, {Urban}, {Urban}, {Verderi},
  {Walkden}, {Wander}, {Weber}, {Wellisch}, {Wenaus}, {Williams}, {Wright},
  {Yamada}, {Yoshida}, {Zschiesche}, \& {G EANT4 Collaboration}}]{geant4}
{Agostinelli}, S., {Allison}, J., {Amako}, K., {et~al.} 2003, Nuclear
  Instruments and Methods in Physics Research A, 506, 250

\bibitem[{{Aharonian} {et~al.}(2005){Aharonian}, {Akhperjanian}, {Aye},
  {Bazer-Bachi}, {Beilicke}, {Benbow}, {Berge}, {Berghaus}, {Bernl{\"o}hr},
  {Boisson}, {Bolz}, {Borgmeier}, {Braun}, {Breitling}, {Brown}, {Gordo},
  {Chadwick}, {Chounet}, {Cornils}, {Costamante}, {Degrange},
  {Djannati-Ata{\"i}}, {Drury}, {Dubus}, {Ergin}, {Espigat}, {Feinstein},
  {Fleury}, {Fontaine}, {Funk}, {Gallant}, {Giebels}, {Gillessen}, {Goret},
  {Hadjichristidis}, {Hauser}, {Heinzelmann}, {Henri}, {Hermann}, {Hinton},
  {Hofmann}, {Holleran}, {Horns}, {de Jager}, {Jung}, {Kh{\'e}lifi}, {Komin},
  {Konopelko}, {Latham}, {Le Gallou}, {Lemi{\`e}re}, {Lemoine}, {Leroy},
  {Lohse}, {Marcowith}, {Masterson}, {McComb}, {de Naurois}, {Nolan},
  {Noutsos}, {Orford}, {Osborne}, {Ouchrif}, {Panter}, {Pelletier}, {Pita},
  {P{\"u}hlhofer}, {Punch}, {Raubenheimer}, {Raue}, {Raux}, {Rayner},
  {Redondo}, {Reimer}, {Reimer}, {Ripken}, {Rob}, {Rolland}, {Rowell},
  {Sahakian}, {Saug{\'e}}, {Schlenker}, {Schlickeiser}, {Schuster}, {Schwanke},
  {Siewert}, {Sol}, {Steenkamp}, {Stegmann}, {Tavernet}, {Terrier},
  {Th{\'e}oret}, {Tluczykont}, {van der Walt}, {Vasileiadis}, {Venter},
  {Vincent}, {Visser}, {V{\"o}lk}, \& {Wagner}}]{Aharonian-05}
{Aharonian}, F., {Akhperjanian}, A.~G., {Aye}, K.-M., {et~al.} 2005, Science,
  307, 1938

\bibitem[{{Anderson} {et~al.}(2014){Anderson}, {Bania}, {Balser}, {Cunningham},
  {Wenger}, {Johnstone}, \& {Armentrout}}]{Anderson-14}
{Anderson}, L.~D., {Bania}, T.~M., {Balser}, D.~S., {et~al.} 2014, \apjs, 212,
  1

\bibitem[{{Bertsch} {et~al.}(1993){Bertsch}, {Dame}, {Fichtel}, {Hunter},
  {Sreekumar}, {Stacy}, \& {Thaddeus}}]{Bertsch-93}
{Bertsch}, D.~L., {Dame}, T.~M., {Fichtel}, C.~E., {et~al.} 1993, \apj, 416,
  587

\bibitem[{{Brown} {et~al.}(2014){Brown}, {Dickey}, {Dawson}, \&
  {McClure-Griffiths}}]{Brown-14}
{Brown}, C., {Dickey}, J.~M., {Dawson}, J.~R., \& {McClure-Griffiths}, N.~M.
  2014, \apjs, 211, 29

\bibitem[{{Carey} {et~al.}(2009){Carey}, {Noriega-Crespo}, {Mizuno}, {Shenoy},
  {Paladini}, {Kraemer}, {Price}, {Flagey}, {Ryan}, {Ingalls}, {Kuchar},
  {Pinheiro Gon{\c c}alves}, {Indebetouw}, {Billot}, {Marleau}, {Padgett},
  {Rebull}, {Bressert}, {Ali}, {Molinari}, {Martin}, {Berriman}, {Boulanger},
  {Latter}, {Miville-Deschenes}, {Shipman}, \& {Testi}}]{Carey-09}
{Carey}, S.~J., {Noriega-Crespo}, A., {Mizuno}, D.~R., {et~al.} 2009, \pasp,
  121, 76

\bibitem[{{Castelletti} {et~al.}(2011){Castelletti}, {Giacani}, {Dubner},
  {Joshi}, {Rao}, \& {Terrier}}]{Castelletti-11}
{Castelletti}, G., {Giacani}, E., {Dubner}, G., {et~al.} 2011, \aap, 536, A98

\bibitem[{{Contreras} {et~al.}(2013){Contreras}, {Schuller}, {Urquhart},
  {Csengeri}, {Wyrowski}, {Beuther}, {Bontemps}, {Bronfman}, {Henning},
  {Menten}, {Schilke}, {Walmsley}, {Wienen}, {Tackenberg}, \&
  {Linz}}]{Contreras-13}
{Contreras}, Y., {Schuller}, F., {Urquhart}, J.~S., {et~al.} 2013, \aap, 549,
  A45

\bibitem[{{Dame} {et~al.}(2001){Dame}, {Hartmann}, \& {Thaddeus}}]{Dame-01}
{Dame}, T.~M., {Hartmann}, D., \& {Thaddeus}, P. 2001, \apj, 547, 792

\bibitem[{{d'Enterria} {et~al.}(2011){d'Enterria}, {Engel}, {Pierog},
  {Ostapchenko}, \& {Werner}}]{Enterria11}
{d'Enterria}, D., {Engel}, R., {Pierog}, T., {Ostapchenko}, S., \& {Werner}, K.
  2011, Astroparticle Physics, 35, 98

\bibitem[{{Dickey} \& {Lockman}(1990)}]{Dickey-90}
{Dickey}, J.~M. \& {Lockman}, F.~J. 1990, \araa, 28, 215

\bibitem[{{Fich} {et~al.}(1989){Fich}, {Blitz}, \& {Stark}}]{Fich-89}
{Fich}, M., {Blitz}, L., \& {Stark}, A.~A. 1989, \apj, 342, 272

\bibitem[{{Fletcher} {et~al.}(1994){Fletcher}, {Gaisser}, {Lipari}, \&
  {Stanev}}]{sibyll}
{Fletcher}, R.~S., {Gaisser}, T.~K., {Lipari}, P., \& {Stanev}, T. 1994, \prd,
  50, 5710

\bibitem[{{Funk} {et~al.}(2007){Funk}, {Hinton}, {P{\"u}hlhofer}, {Aharonian},
  {Hofmann}, {Reimer}, \& {Wagner}}]{Funk-07}
{Funk}, S., {Hinton}, J.~A., {P{\"u}hlhofer}, G., {et~al.} 2007, \apj, 662, 517

\bibitem[{{Giacani} {et~al.}(2008){Giacani}, {Castelletti}, {Dubner}, {Joshi},
  {Rao}, \& {Terrier}}]{Giacani-08}
{Giacani}, E., {Castelletti}, G., {Dubner}, G., {et~al.} 2008, in American
  Institute of Physics Conference Series, Vol. 1085, American Institute of
  Physics Conference Series, ed. F.~A. {Aharonian}, W.~{Hofmann}, \&
  F.~{Rieger}, 234--236

\bibitem[{{Gotthelf} {et~al.}(2014){Gotthelf}, {Tomsick}, {Halpern}, {Gelfand},
  {Harrison}, {Boggs}, {Christensen}, {Craig}, {Hailey}, {Kaspi}, {Stern}, \&
  {Zhang}}]{Gotthelf-14}
{Gotthelf}, E.~V., {Tomsick}, J.~A., {Halpern}, J.~P., {et~al.} 2014, \apj,
  788, 155

\bibitem[{{Griffin} {et~al.}(2010){Griffin}, {Abergel}, {Abreu}, {Ade},
  {Andr{\'e}}, {Augueres}, {Babbedge}, {Bae}, {Baillie}, {Baluteau}, {Barlow},
  {Bendo}, {Benielli}, {Bock}, {Bonhomme}, {Brisbin}, {Brockley-Blatt},
  {Caldwell}, {Cara}, {Castro-Rodriguez}, {Cerulli}, {Chanial}, {Chen},
  {Clark}, {Clements}, {Clerc}, {Coker}, {Communal}, {Conversi}, {Cox},
  {Crumb}, {Cunningham}, {Daly}, {Davis}, {de Antoni}, {Delderfield}, {Devin},
  {di Giorgio}, {Didschuns}, {Dohlen}, {Donati}, {Dowell}, {Dowell}, {Duband},
  {Dumaye}, {Emery}, {Ferlet}, {Ferrand}, {Fontignie}, {Fox}, {Franceschini},
  {Frerking}, {Fulton}, {Garcia}, {Gastaud}, {Gear}, {Glenn}, {Goizel},
  {Griffin}, {Grundy}, {Guest}, {Guillemet}, {Hargrave}, {Harwit}, {Hastings},
  {Hatziminaoglou}, {Herman}, {Hinde}, {Hristov}, {Huang}, {Imhof}, {Isaak},
  {Israelsson}, {Ivison}, {Jennings}, {Kiernan}, {King}, {Lange}, {Latter},
  {Laurent}, {Laurent}, {Leeks}, {Lellouch}, {Levenson}, {Li}, {Li},
  {Lilienthal}, {Lim}, {Liu}, {Lu}, {Madden}, {Mainetti}, {Marliani}, {McKay},
  {Mercier}, {Molinari}, {Morris}, {Moseley}, {Mulder}, {Mur}, {Naylor},
  {Nguyen}, {O'Halloran}, {Oliver}, {Olofsson}, {Olofsson}, {Orfei}, {Page},
  {Pain}, {Panuzzo}, {Papageorgiou}, {Parks}, {Parr-Burman}, {Pearce},
  {Pearson}, {P{\'e}rez-Fournon}, {Pinsard}, {Pisano}, {Podosek}, {Pohlen},
  {Polehampton}, {Pouliquen}, {Rigopoulou}, {Rizzo}, {Roseboom}, {Roussel},
  {Rowan-Robinson}, {Rownd}, {Saraceno}, {Sauvage}, {Savage}, {Savini},
  {Sawyer}, {Scharmberg}, {Schmitt}, {Schneider}, {Schulz}, {Schwartz},
  {Shafer}, {Shupe}, {Sibthorpe}, {Sidher}, {Smith}, {Smith}, {Smith},
  {Spencer}, {Stobie}, {Sudiwala}, {Sukhatme}, {Surace}, {Stevens}, {Swinyard},
  {Trichas}, {Tourette}, {Triou}, {Tseng}, {Tucker}, {Turner}, {Vaccari},
  {Valtchanov}, {Vigroux}, {Virique}, {Voellmer}, {Walker}, {Ward}, {Waskett},
  {Weilert}, {Wesson}, {White}, {Whitehouse}, {Wilson}, {Winter}, {Woodcraft},
  {Wright}, {Xu}, {Zavagno}, {Zemcov}, {Zhang}, \& {Zonca}}]{Griffin-10}
{Griffin}, M.~J., {Abergel}, A., {Abreu}, A., {et~al.} 2010, \aap, 518, L3

\bibitem[{{Kafexhiu} {et~al.}(2014){Kafexhiu}, {Aharonian}, {Taylor}, \&
  {Vila}}]{Kafexhiu-14}
{Kafexhiu}, E., {Aharonian}, F., {Taylor}, A.~M., \& {Vila}, G.~S. 2014, \prd,
  90, 123014

\bibitem[{{Kalmykov} {et~al.}(1997){Kalmykov}, {Ostapchenko}, \&
  {Pavlov}}]{qg01}
{Kalmykov}, N.~N., {Ostapchenko}, S.~S., \& {Pavlov}, A.~I. 1997, Nuclear
  Physics B Proceedings Supplements, 52, 17

\bibitem[{{Kamae} {et~al.}(2006){Kamae}, {Karlsson}, {Mizuno}, {Abe}, \&
  {Koi}}]{Kamae06}
{Kamae}, T., {Karlsson}, N., {Mizuno}, T., {Abe}, T., \& {Koi}, T. 2006, \apj,
  647, 692

\bibitem[{{Kelner} {et~al.}(2006){Kelner}, {Aharonian}, \&
  {Bugayov}}]{Kelner06}
{Kelner}, S.~R., {Aharonian}, F.~A., \& {Bugayov}, V.~V. 2006, \prd, 74, 034018

\bibitem[{{Lemiere} {et~al.}(2009){Lemiere}, {Slane}, {Gaensler}, \&
  {Murray}}]{Lemiere-09}
{Lemiere}, A., {Slane}, P., {Gaensler}, B.~M., \& {Murray}, S. 2009, \apj, 706,
  1269

\bibitem[{{Lemoine-Goumard} {et~al.}(2014){Lemoine-Goumard}, {Grondin},
  {Acero}, {Ballet}, {Laffon}, \& {Reposeur}}]{Lemoine-Goumard-14}
{Lemoine-Goumard}, M., {Grondin}, M.-H., {Acero}, F., {et~al.} 2014, \apjl,
  794, L16

\bibitem[{{McClure-Griffiths} {et~al.}(2005){McClure-Griffiths}, {Dickey},
  {Gaensler}, {Green}, {Haverkorn}, \& {Strasser}}]{McClure-Griffiths-05}
{McClure-Griffiths}, N.~M., {Dickey}, J.~M., {Gaensler}, B.~M., {et~al.} 2005,
  \apjs, 158, 178

\bibitem[{{Poglitsch} {et~al.}(2010){Poglitsch}, {Waelkens}, {Geis},
  {Feuchtgruber}, {Vandenbussche}, {Rodriguez}, {Krause}, {Renotte}, {van
  Hoof}, {Saraceno}, {Cepa}, {Kerschbaum}, {Agn{\`e}se}, {Ali}, {Altieri},
  {Andreani}, {Augueres}, {Balog}, {Barl}, {Bauer}, {Belbachir}, {Benedettini},
  {Billot}, {Boulade}, {Bischof}, {Blommaert}, {Callut}, {Cara}, {Cerulli},
  {Cesarsky}, {Contursi}, {Creten}, {De Meester}, {Doublier}, {Doumayrou},
  {Duband}, {Exter}, {Genzel}, {Gillis}, {Gr{\"o}zinger}, {Henning},
  {Herreros}, {Huygen}, {Inguscio}, {Jakob}, {Jamar}, {Jean}, {de Jong},
  {Katterloher}, {Kiss}, {Klaas}, {Lemke}, {Lutz}, {Madden}, {Marquet},
  {Martignac}, {Mazy}, {Merken}, {Montfort}, {Morbidelli}, {M{\"u}ller},
  {Nielbock}, {Okumura}, {Orfei}, {Ottensamer}, {Pezzuto}, {Popesso},
  {Putzeys}, {Regibo}, {Reveret}, {Royer}, {Sauvage}, {Schreiber}, {Stegmaier},
  {Schmitt}, {Schubert}, {Sturm}, {Thiel}, {Tofani}, {Vavrek}, {Wetzstein},
  {Wieprecht}, \& {Wiezorrek}}]{Poglitsch-10}
{Poglitsch}, A., {Waelkens}, C., {Geis}, N., {et~al.} 2010, \aap, 518, L2

\bibitem[{{Rosolowsky} {et~al.}(2010){Rosolowsky}, {Dunham}, {Ginsburg},
  {Bradley}, {Aguirre}, {Bally}, {Battersby}, {Cyganowski}, {Dowell},
  {Drosback}, {Evans}, {Glenn}, {Harvey}, {Stringfellow}, {Walawender}, \&
  {Williams}}]{Rosolowsky-10}
{Rosolowsky}, E., {Dunham}, M.~K., {Ginsburg}, A., {et~al.} 2010, \apjs, 188,
  123

\bibitem[{{Schuller} {et~al.}(2009){Schuller}, {Menten}, {Contreras},
  {Wyrowski}, {Schilke}, {Bronfman}, {Henning}, {Walmsley}, {Beuther},
  {Bontemps}, {Cesaroni}, {Deharveng}, {Garay}, {Herpin}, {Lefloch}, {Linz},
  {Mardones}, {Minier}, {Molinari}, {Motte}, {Nyman}, {Reveret}, {Risacher},
  {Russeil}, {Schneider}, {Testi}, {Troost}, {Vasyunina}, {Wienen}, {Zavagno},
  {Kovacs}, {Kreysa}, {Siringo}, \& {Wei{\ss}}}]{Schuller-09}
{Schuller}, F., {Menten}, K.~M., {Contreras}, Y., {et~al.} 2009, \aap, 504, 415

\bibitem[{{Sj{\"o}strand} {et~al.}(2008){Sj{\"o}strand}, {Mrenna}, \&
  {Skands}}]{pythia}
{Sj{\"o}strand}, T., {Mrenna}, S., \& {Skands}, P. 2008, Computer Physics
  Communications, 178, 852

\bibitem[{{Slane} {et~al.}(2010){Slane}, {Castro}, {Funk}, {Uchiyama},
  {Lemiere}, {Gelfand}, \& {Lemoine-Goumard}}]{Slane-10}
{Slane}, P., {Castro}, D., {Funk}, S., {et~al.} 2010, \apj, 720, 266

\bibitem[{{Urquhart} {et~al.}(2012){Urquhart}, {Hoare}, {Lumsden}, {Oudmaijer},
  {Moore}, {Mottram}, {Cooper}, {Mottram}, \& {Rogers}}]{Urquhart-12}
{Urquhart}, J.~S., {Hoare}, M.~G., {Lumsden}, S.~L., {et~al.} 2012, \mnras,
  420, 1656

\bibitem[{{Werner}(1993)}]{Werner93}
{Werner}, K. 1993, \physrep, 232, 87

\bibitem[{{Whiteoak} \& {Green}(1996)}]{Whiteoak-96}
{Whiteoak}, J.~B.~Z. \& {Green}, A.~J. 1996, \aaps, 118, 329

\end{thebibliography}

\IfFileExists{\jobname.bbl}{}
{\typeout{}
\typeout{****************************************************}
\typeout{****************************************************}
\typeout{** Please run "bibtex \jobname" to optain}
\typeout{** the bibliography and then re-run LaTeX}
\typeout{** twice to fix the references!}
\typeout{****************************************************}
\typeout{****************************************************}
\typeout{}
}

\end{document}